\documentclass[cits]{PoS}
\pdfoutput=1
\usepackage{multicol}
\usepackage{slashed}
\usepackage{setspace}
\usepackage{amsmath}
\usepackage{bm}
\usepackage{color}
\usepackage{booktabs}
\usepackage{epstopdf}
\usepackage[pdftex]{graphicx}
\usepackage[symbol*,hang,flushmargin]{footmisc} %Circumvents mathptmx package that is auto loaded in PoS.
\DeclareMathAlphabet{\mathcal}{OMS}{cmsy}{m}{n}

\definecolor{gray}{rgb}{0.8,0.8,0.8}
\newcommand{\ccc}[1]{} % Jason
\newcommand{\ask}[1]{\textcolor{blue}{#1}} % Andreas
\title{Update on a short-distance $\bm{D^0}$-meson mixing calculation with $\bm{N_f=2+1}$ flavors}

\ShortTitle{$D^0$-meson mixing update}

\author{\speaker{C.C. Chang}\textsuperscript{\textnormal{\textit{a, f}}}, 
C. Bernard\textsuperscript{\textnormal{\textit{b}}}, 
C.M. Bouchard\textsuperscript{\textnormal{\textit{c}}}, 
A.X. El-Khadra\textsuperscript{\textnormal{\textit{a, f}}}, 
E.D. Freeland\textsuperscript{\textnormal{\textit{d}}}, 
E.~G\'amiz\textsuperscript{\textnormal{\textit{e}}}, 
A.S. Kronfeld\textsuperscript{\textnormal{\textit{f, g}}}, 
J. Laiho\textsuperscript{\textnormal{\textit{h}}}, 
R.S. Van de Water\textsuperscript{\textnormal{\textit{f}}}\\
    \textsuperscript{a}Physics Department, University of Illinois, Urbana, IL 61801, USA\\
    \textsuperscript{b}Department of Physics, Washington University, St. Louis, MO 63130, USA\\
    \textsuperscript{c}Department of Physics, The Ohio State University, Columbus, OH 43210, USA\\
    \textsuperscript{d}Liberal Arts Department, The School of the Art Institute of Chicago, Chicago, 
        IL 60603, USA\\
    \textsuperscript{e}CAFPE and Departamento de Fisica Teorica y del Cosmos, Universidad de Granada, 
        E-18002 Granada, Spain\\
    \textsuperscript{f}Theoretical Physics Department, Fermi National Accelerator Laboratory,\thanks{Operated
        by Fermi Research Alliance, LLC, under Contract No.\ DE-AC02-07CH11359 with the United 
        States Department of Energy} { } Batavia, IL 60510, USA\\ 
    \textsuperscript{g}Institute for Advanced Study, Technische Universit\"at M\"unchen,
        85748 Garching, Germany\\
    \textsuperscript{h}Department of Physics, Syracuse University, Syracuse, NY 13244, USA\\}
\author{Fermilab Lattice and MILC Collaborations\\
        E-mail: \email{cchang5@illinois.edu}}

\abstract{We present an update on our calculation of the short-distance $D^0$-meson mixing hadronic matrix
elements.
The analysis is performed on the MILC collaboration's $N_f=2+1$ asqtad configurations.
We use asqtad light valence quarks and the Sheikoleslami-Wohlert action with the Fermilab interpretation for
the valence charm quark.
SU(3), partially quenched, rooted, staggered heavy-meson chiral perturbation theory is used to extrapolate
to the chiral-continuum limit.
Systematic errors arising from the chiral-continuum extrapolation, heavy-quark discretization, and
quark-mass uncertainties are folded into the statistical errors from the chiral-continuum fits with methods
of Bayesian inference.
A preliminary error budget for all five operators is presented.}

\FullConference{The 32nd International Symposium on Lattice Field Theory,\\
		23-28 June, 2014\\
		Columbia University New York, NY}

\begin{document}
\section{Introduction}
$D^0$-meson mixing is currently the least well understood meson mixing process.
Experimental efforts underway or planned at LHCb, BES III, and Belle
II, should improve our understanding and ignite excitement for the future of charm physics.
In the Standard Model (SM), the short-distance contributions to $D^0$-meson mixing are GIM suppressed by
$m_s^2-m_d^2$ and Cabbibo suppressed by $|V_{ub}V^*_{cb}|^2$; therefore $D^0$-meson mixing is expected to
receive significant contributions from the long-distance processes in the Standard Model.
However, it is also possible for $D^0$-meson mixing to receive enhancements from short-distance new physics
(NP) contributions.
Therefore, in conjunction with next generation flavor factories, knowledge of the five short-distance
hadronic matrix elements will allow for model-discrimination between NP theories~\cite{0705.3650}.
The short-distance matrix elements are described by a basis of five 4-quark operators that are invariant
under Lorentz, Fierz, charge conjugation, parity inversion, and time reversal transformations and may be expressed as the following:
\begin{align}
\mathcal{O}_1=\bar{c}^\alpha \gamma^\mu L u^\alpha \bar{c}^\beta \gamma^\mu Lu^\beta, &&
\mathcal{O}_2=\bar{c}^\alpha L u^\alpha \bar{c}^\beta L u^\beta, &&
\mathcal{O}_3=\bar{c}^\alpha L u^\beta \bar{c}^\beta L u^\alpha,
\label{intro_sm}
\end{align}
\begin{align}
\mathcal{O}_4=\bar{c}^\alpha L u^\alpha \bar{c}^\beta R u^\beta, &&
\mathcal{O}_5=\bar{c}^\alpha L u^\beta \bar{c}^\beta R u^\alpha,
\label{intro_bsm}
\end{align}
where $L$ and $R$ are the left- and right-handed projection operators, while $c$ and $u$ denote the charm
and up quarks respectively.
The operators in Eq.~(\ref{intro_bsm}) couple to right-handed quarks and, therefore, appear only in NP 
scenarios.

\section{Lattice setup and correlator analysis}

The correlators pertinent to this project are constructed on a large subset of the MILC gauge
configurations~\cite{0903.3598} with 2+1 asqtad staggered sea quarks.
A complete list of ensembles used for this project is given in Ref.~\cite{1311.6820}.
The light valence quarks are also generated with the asqtad action, with masses ranging from $m_s$ to
$m_s/20$.
We have a large range of valence masses, hence we use partially quenched chiral perturbation theory to
extrapolate to physical up- and down-quark masses.
For the heavy charm quark, we use the Sheikoleslami-Wohlert action with the Fermilab interpretation, which
ensures that the couplings in the theory are smoothly bounded for $am_q \slashed{\ll} 1$, as well as in the limit $am_q\rightarrow 0$, resulting in well controlled errors.
The heavy-quark Lagrangian is tree-level improved and the lattice operators corresponding to the
$\mathcal{O}_i$ use rotated heavy-quark fields, resulting in errors starting at $\mathrm{O}(\alpha_s a,
a^2)$.

Results of the correlator analysis have been presented in Ref\ask{.}~\cite{1311.6820} and are complete for
all five 4-quark operators.
Under renormalization, the sets of operators given in Eq.~(\ref{intro_sm}) and (\ref{intro_bsm}) mix among
each other.
Thus,
\begin{equation}
\left<D^0|\mathcal{O}_i|\bar{D}^0\right>^{\mathrm{\overline{MS}-NDR}}(m_c)=\sum_{j=1}^5[\delta_{ij}+\alpha_s(q^*)\zeta^{\mathrm{\overline{MS}-NDR}}_{ij}(am_c)+\mathrm{O}(\alpha_s^2(q^*))]\left<D^0|\mathcal{O}_j|\bar{D}^0\right>^\text{lat}.
\label{renorm}
\end{equation}
The $\zeta^{\mathrm{\overline{MS}-NDR}}_{ij}$s are matching coefficients relating one-loop lattice and continuum renormalizations
evaluated at the charm quark mass $m_c$, and $\alpha_s(q^*)$ is the strong coupling discussed in
Ref.~\cite{PhysRevD.28.228}.
The one-loop continuum calculations require choosing an additional set of evanescent operators during
intermediate steps of dimensional regularization.
We report results using the BBGLN~\cite{9808385} scheme.
Once the analysis has been finalized, results in the BJU~\cite{0005183} scheme will also be reported.
The errors from renormalization and matching start at $\mathrm{O}(\alpha_s^2)$, as suggested by
Eq.~(\ref{renorm}).
For brevity, we will use the short-hand 
$\left<\mathcal{O}_i\right>\equiv\left<D^0|\mathcal{O}_i|\bar{D}^0\right>^\mathrm{\overline{MS}-NDR}(m_c)$ 
when referring the renormalized matrix element below.

The charm quark mass is set by tuning the $D_s$-meson mass to its physical 
value~\cite{1003.1937,Bailey:2014tva}.
Corrections to the slight mistunings are implemented by linearly extrapolating the matrix element to the
correct (tuned) charm mass $m_c$,
\begin{align}
\left<\mathcal{O}_i\right>_\text{tune}=&\left<\mathcal{O}_i\right>+\sigma_{i}\Delta\left(1/M_2\right),
\label{kappatune}
\end{align}
where $\sigma_{i}$ is the slope of $\left<\mathcal{O}_i\right>$ with respect to the inverse kinetic mass $1/M_2$ and
is obtained by performing a correlated unconstrained fit on the $a\approx0.12$~fm, $m_l/m_s=0.2$ ensemble at
two valence mass points and two values of $m_c$.
The errors of the corrections to the heavy-quark mistuning are determined by the precision of the linear
fits performed to extract $\sigma_i$ as well as the determination of the tuned $m_c$, outlined in
Refs.~\cite{1003.1937,Bailey:2014tva}.

\section{Chiral-continuum extrapolation}
To extrapolate to the chiral-continuum limit, we use SU(3), partially quenched, rooted, staggered,
heavy-meson chiral perturbation theory~\cite{1303.0435}. For reviews of heavy meson and staggered chiral perturbation theory see:~\cite{Manohar:429724,0912.4042}.
The expression has the schematic form,
\begin{equation}
\left<\mathcal{O}_i\right>=\beta_i\left(1+\frac{\mathcal{W}_{u\bar{c}}+\mathcal{W}_{c\bar{u}}}{2}+\mathcal{T}_u^{(i)}+\frac{C(\beta_{j\neq i})}{\beta_i}\tilde{\mathcal{T}}_u^{(i)}+\text{analytic terms}\right)+\beta'_i\left(\mathcal{Q}_u^{(i)}+\frac{D(\beta'_{j\neq i})}{\beta'_i}\tilde{\mathcal{Q}}^{(i)}_u\right).
\label{chiptexpression}
\end{equation}
The $\beta$s and $\beta'$s along with the coefficents in the analytic terms are the low energy constants
(LECs) of the theory and are determined from fits to the matrix element data.
The functions $C$ and $D$ are linear in $\beta_{j\neq i}$ for each $\left<\mathcal{O}_i\right>$,
introducing mixing between the leading-order LECs within the sets
$\{\left<\mathcal{O}_1\right>,\left<\mathcal{O}_2\right>,\left<\mathcal{O}_3\right>\}$ and
$\{\left<\mathcal{O}_4\right>,\left<\mathcal{O}_5\right>\}$.
The terms $\mathcal{W}$, $\mathcal{T}$ and $\mathcal{Q}$ denote the chiral logarithms arising from the
wavefunction renormalization, tadpole, and sunset one-loop Feynman diagrams.
Using staggered light quarks and local (not point-split) operators introduces wrong-spin taste-mixing chiral logarithms $\tilde{\mathcal{T}}$ and
$\tilde{\mathcal{Q}}$, however these contributions do not introduce new LECs, as indicated by the $C$ and $D$
functions in Eq.~(\ref{chiptexpression}).
%The origin of mixing stems from transforming from the spin-taste basis to the SUSY basis of operators listed in Eq.~(\ref{intro_sm}, \ref{intro_bsm}), introducing no new LECs.
From examining the correlator fits, we observe the $\beta$s to be of order~1.
The $\beta$s are introduced into the fit via priors and are loosely constrained with a prior width of 10,
such that the data determine their values.

Analytic terms in the chiral fit capture the effects of explicit NLO SU(3)
flavor symmetry breaking and SU(4) taste breaking as well as NNLO contributions.
The dependence of the valence mass, sea mass and taste breaking effects are parameterized by dimensionless
``natural $\chi$PT'' parameters, which yield coefficients that are naturally of
$\mathrm{O}(1)$~\cite{1112.3051},
\begin{align}
x_{u,l,s}\equiv \frac{(r_1B_0)(r_1/a)(2am_{u,l,s})}{8\pi^2f_\pi^2r_1^2} && x_{\bar{\Delta}}\equiv \frac{r_1^2a^2\bar{\Delta}}{8\pi^2f_\pi^2r_1^2}
\end{align}
where $m_{u,l,s}$ corresponds to the valence up, sea up/down and sea strange masses and $\bar{\Delta}$ is
the average taste splitting.
The NLO and NNLO analytic terms are,
\begin{align}
\left.\text{NLO analytic}\right.=&c_0 x_u + c_1 (2x_l + x_s) + c_2 x_{\bar{\Delta}}\\
\left.\text{NNLO analytic}\right.=&\sum_j d_j F_j(x_nx_m)
\end{align}
where the NLO coefficients $c_i$ are loosely constrained while the NNLO coefficients $d_j$ are constrained
to be $\mathrm{O}(1)$.
The functions $F_j(x_nx_m)$ represents the set of quadratic functions in $x_{u,l,s,\bar{\Delta}}$.
Our fits results are insensitive to the addition of terms beyond NNLO.

The largest $1/M_D$ corrections from heavy-meson $\chi$PT arise from the spin splittings (e.g., $M_{D^*}-M_D$) and flavor splittings (e.g., $M_{D_s}-M_D$) and are accounted for in the chiral logarithms presented in Eq.~(\ref{chiptexpression}).

For LECs that cannot be determined by the data, priors are used to constrain the parameters.
The largest parametric uncertainty that enters the chiral-continuum extrapolation is the heavy-meson
coupling and is accounted for through the prior $g_{D^*D\pi}=0.53\pm 0.08$~\cite{1210.5410}.
Other parameters such as the hyperfine splitting $\Delta^*$ is determined
by experimental results~\cite{pdg} and are introduced as priors to incorporate experimental uncertainties.
Systematic errors arising from the free parameters of the effective theory are all accounted for in the
chiral fits.

Along with the chiral-continuum extrapolation, we fold in the $\mathrm{O}(\alpha_s a,a^2)$ heavy-quark
discretization errors arising from the Lagrangian and operator.
We estimate the contributions arising from
perturbative matching between continuum QCD and lattice QCD through HQET~\cite{0911.5432,Kronfeld:2000ck},
\begin{align}
\mathcal{L}_\text{QCD}\doteq &\mathcal{L}_\text{HQET}=\sum_k C_k^\text{cont}(m_c)\mathcal{O}_k,\\
\mathcal{L}_\text{lat}\doteq &\mathcal{L}_{\text{HQET}(m_0a)}= \sum_k C_k^\text{lat}(m_c,m_0a)\mathcal{O}_k,
\end{align}
from which it follows that the error from each term is
\begin{align}
\text{error}_k=&\left|\left[C_k^\text{lat}(m_c,m_0a)-C^\text{cont}_k(m_c)\right]\left<\mathcal{O}_k\right>\right|.
\end{align}
The $\mathrm{O}(a\alpha_s, a^2)$ discretization effects are included as part of the chiral-continuum fit.
 %as,
%\begin{align}
%\left.\text{error}_i\right.=&\beta_i\sum_n z_n f_n(m_0a).
%\label{HQerror}
%\text{error}_i\equiv \left<\mathcal{O}_i\right>_\text{HQ error}&= \beta_i \left\{(a\Lambda_\text{HQET})z_{B7}f_{B7}(m_0a)\right.\nonumber\\
%&+(a\Lambda_\text{HQET})^2\left[z_Ef_E(m_0a)+z_Xf_X(m_0a)+z_Yf_Y(m_0a)\right]\nonumber\\
%&\left.+(a\Lambda_\text{HQET})^3z_2f_2(m0a)\right\}.
%\end{align}
The mismatch functions are discussed in detail in Ref.~\cite{1112.3051}. Their coefficients introduced as priors with central values of 0, and widths determined by power counting and are $\mathrm{O}(1)$.

\section{Error budget}

\begin{figure}
	\centering
		\includegraphics[width=1.00\textwidth]{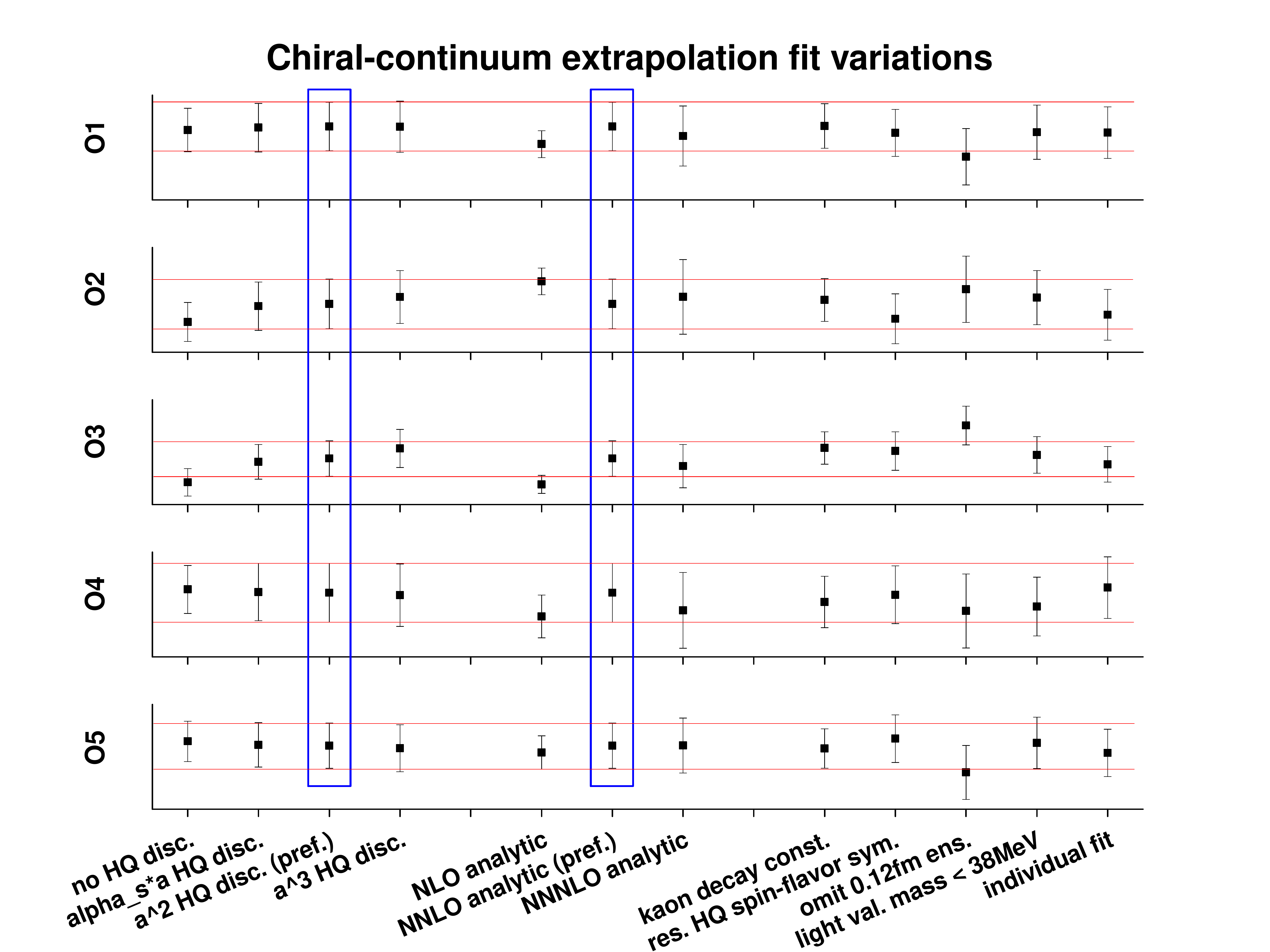}
	\caption{Fit variations for the chiral-continuum extrapolation. The preferred fit is indicated by the blue 
        boxes, with $1\sigma$ error bands from that fit indicated by the horizontal red lines. The first group of fits 
        explores options for including heavy-quark discretization errors. The second group explores adding 
        chiral analytic terms. The last group explores parameter and data set changes.}
	\label{fig:chiptfitvar}
\end{figure}

The stability of the chiral-continuum extrapolation is demonstrated in Fig.~\ref{fig:chiptfitvar}.
The preferred fit (blue boxes) is a simultaneous fit over the operators that mix in the $\chi$PT over four
lattice spacings, including $\mathrm{O}(a\alpha_s, a^2)$ heavy-quark discretization corrections and NNLO
analytic terms.
The first group of four fits in Fig.~\ref{fig:chiptfitvar} progressively introduces heavy-quark
discretization terms to the fit.
The second group of three fits progressively includes more chiral analytic terms.
In both cases, the preferred fit lies in the region of stability. The largest changes in the central value occur when introducing the $\mathrm{O}(\alpha_s a)$ heavy quark discretization errors, and NNLO analytic terms, showing that truncation errors of the respective expansions are included as part of the preferred fit.

The third set of fits differ from the preferred fits as follows: changing $f_\pi$ to $f_K$, restoring
heavy quark flavor-spin symmetry, omitting the 0.12~fm ensembles, omitting larger $(>38\text{MeV})$ light valence quark masses, and
performing fits individually for each operator.
The fits are stable within one standard deviation.

By incorporating the heavy-quark discretization errors into the chiral-continuum fit, the associated relative error
ranges from 2.3--2.9\%.
%We may also estimate this error through power counting by evaluating the mismatch functions on the 0.045 fm
%ensemble, combined with coefficients set to $1$.
%This suggests an error of 2.1\%.
%The errors determined by both methods are consistent.
We estimate the renormalization error by setting the $\alpha_s^2$ two-loop coefficients to~1 and using an average value of $\alpha_s^2$ across all four lattice spacing. This suggests a systematic
error of 6.5\%. Based on previous $D$-meson decay constant analysis~\cite{1112.3051}, we expect the finite volume effects to be $<1\%$.

\begin{table}[htbp]
	\centering
		\begin{tabular}{lccccc}
		\toprule
		\toprule
		& $\left<\mathcal{O}_1\right>$	& $\left<\mathcal{O}_2\right>$	& $\left<\mathcal{O}_3\right>$	& $\left<\mathcal{O}_4\right>$	& $\left<\mathcal{O}_5\right>$\\
		\midrule
		Statistical & 3.2\% & 2.1\% & 3.3\% & 2.2\% & 3.9\%\\
		Chiral extrapolation & 2.2\% & 2.1\% & 2.4\% & 2.1\% & 3.0\%\\
		$(\alpha_s a, a^2)$ HQ error & 2.4\% & 2.4\% & 2.9\% & 2.3\% & 2.7\%\\
		HQ mass tuning & 0.8\% & 1.1\% & 1.0\% & 1.2\% & 1.3\%\\
		\cmidrule{2-6}
		Renormalization & \multicolumn{5}{c}{6.5\%} \\ %5.5\% & 3.6\% & 2.7\% & 1 & 1\\
    Finite volume & \multicolumn{5}{c}{$<1\%$}\\
		\midrule
		Total error & 8.0\% & 7.7\% & 8.3\% & 7.7\% & 8.7\%\\
		\bottomrule
		\bottomrule
		\end{tabular}
	\caption{Preliminary error budget for $D^0$-meson mixing hadronic matrix elements in the continuum and for physical quark masses.  Values are percent relative errors.}
	\label{errortable}
\end{table}

\section{Conclusions and outlook}
Our chiral-continuum analysis of the $D^0$-meson hadronic matrix elements, including a complete error budget, is near completion.
Due to the large contribution of the renormalization error, a partially nonperturbative approach to
determining the renormalization coefficients is currently being investigated.
The results of the matrix elements will also be combined with our decay constants calculated
separately~\cite{1403.6796,1407.3772} and bag parameters will be reported.

\section*{Acknowledgements}
This work was supported by the U.S.\ Department of Energy, the National Science Foundation, the Universities
Research Association, the MINECO, Junta de Andaluc\'ia, the European Commission, the German Excellence
Initiative, the European Union Seventh Framework Programme, and the European Union's Marie Curie COFUND
program.
Computation for this work was done at the Argonne Leadership Computing Facility (ALCF), the National Center
for Atmospheric Research (UCAR), the National Center for Supercomputing Resources (NCSA), the National
Energy Resources Supercomputing Center (NERSC), the National Institute for Computational Sciences (NICS),
the Texas Advanced Computing Center (TACC), and the USQCD facilities at Fermilab, under grants from the NSF
and DOE.


\begin{thebibliography}{99}

%\cite{0705.3650}
\bibitem{0705.3650} 
  E.~Golowich, J.~Hewett, S.~Pakvasa and A.~A.~Petrov,
  %``Implications of $D^0$ - $\bar{D}^0$ Mixing for New Physics,''
  Phys.\ Rev.\ D{\bf 76}, 095009 (2007)
  [arXiv:0705.3650 [hep-ph]].
  %%CITATION = ARXIV:0705.3650;%%
  %162 citations counted in INSPIRE as of 24 Nov 2013

%\cite{0903.3598}
\bibitem{0903.3598} 
  A.~Bazavov {\it et al.},
  %``Nonperturbative QCD simulations with 2+1 flavors of improved staggered quarks,''
  Rev.\ Mod.\ Phys.\  {\bf 82}, 1349 (2010)
  [arXiv:0903.3598 [hep-lat]].
  %%CITATION = ARXIV:0903.3598;%%
  %217 citations counted in INSPIRE as of 24 Nov 2013

\bibitem{1311.6820}
      C.~C.~Chang {\it et al.} [Fermilab Lattice and MILC Collaborations],
			\pos{PoS(Lattice 2013)477} (2013)
			[arXiv:1311.6820 [hep-lat]].
			
\bibitem{0911.5432} 
  R.~T.~Evans {\it et al.}  [Fermilab Lattice and MILC Collaborations],
  %``B-Bbar Mixing and Matching with Fermilab Heavy Quarks,''
  %PoS LAT {\bf 2009}, 245 (2009),
	\pos{PoS(LAT2009)245} (2009)
	[arXiv:0911.5432 [hep-lat]].
  %%CITATION = ARXIV:0911.5432;%%
  %11 citations counted in INSPIRE as of 24 Nov 2013

\bibitem{PhysRevD.28.228}
  S.~J.~Brodsky, G.~P.~Lepage and P.~B.~Mackenzie,
  Phys.\ Rev.\ D{\bf 28}, 228--235 (1983)


\bibitem{9808385}
  M.~Beneke {\it et al.},
	Phys.\ Lett.\ B{\bf 459}, 631-640 (1999)
	[hep-ph/9808385].

\bibitem{0005183}
A.~J.~Buras, M.~Misiak, and J.~Urban,
  %``Two loop QCD anomalous dimensions of flavor changing four quark operators within and beyond the standard model,''
  Nucl.\ Phys.\ B {\bf 586}, 397 (2000)
  [hep-ph/0005183].
  %%CITATION = HEP-PH/0005183;%%

\bibitem{1003.1937}
      C.~Bernard {\it et al.} [Fermilab Lattice and MILC Collaborations],
      Phys.\ Rev.\ D {\bf 83}, 034503 (2011)
      [arXiv:1003.1937 [hep-lat]].
%
\bibitem{Bailey:2014tva} 
  J.~A.~Bailey {\it et al.} [Fermilab Lattice and MILC Collaborations],
  %``Update of $V_{cb}$ from the $\bar{B}\to D^*\ell\bar{\nu}$ form factor at zero recoil with three-flavor lattice QCD,''
  Phys.\ Rev.\ D {\bf 89}, 114504 (2014)
  [arXiv:1403.0635 [hep-lat]].
  %%CITATION = ARXIV:1403.0635;%%
  %8 citations counted in INSPIRE as of 11 Nov 2014
  
\bibitem{0912.4042}
      M.~Golterman,
			arXiv:0912.4042 [hep-lat].

\bibitem{Manohar:429724}
      A.~V.~Manohar and M.~B.~Wise,
      \emph{Heavy Quark Physics},
      Cambridge Univ. Press (2000).

\bibitem{1303.0435}
      C.~Bernard [MILC Collaboration],
			Phys.\ Rev.\ D {\bf 87}, 114503 (2013)
      [arXiv:1303.0435 [hep-lat]].
			
\bibitem{1112.3051}
      A.~Bazavov {\it et al.} [Fermilab Lattice and MILC Collaborations],
      Phys.\ Rev.\ D {\bf 85}, 114506 (2012)
			[arXiv:1112.3051 [hep-lat]]

\bibitem{1210.5410}
			D.~Becirevic and F.~Sanfilippo,
			Phys.\ Lett.\ B {\bf 721}, 94-100 (2013)
			[arXiv:1210.5410 [hep-lat]].
			
%\bibitem{1002.1655}
			%J.~L.~Rosner and S.~Stone,
			%[arXiv:1002.1655 [hep-ex]].
%
%\bibitem{Beringer:1900zz}
      %J.~Beringer {\it et al.} [Particle Data Group],
			%Phys.\ Rev.\ D {\bf 86}, 010001 (2012).

\bibitem{pdg}
	K.~A.~Olive {\it et al.} [Particle Data Group],
	Chin.\ Phys.\ C, {\bf 38}, 090001 (2014).

\bibitem{Kronfeld:2000ck} 
  A.~S.~Kronfeld,
  %``Application of heavy quark effective theory to lattice QCD. 1. Power corrections,''
  Phys.\ Rev.\ D {\bf 62}, 014505 (2000)
  [hep-lat/0002008].
  %%CITATION = HEP-LAT/0002008;%%
  %79 citations counted in INSPIRE as of 04 Nov 2014

%\bibitem{0910.1229}
			%C.~T.~H.~Davies {\it et al.} [HPQCD Collaboration],
			%Phys.\ Rev.\ D {\bf 81}, 034506 (2010)
			%[arXiv: 0910.1229 [hep-lat]].
%
%\bibitem{LSQFit}
			%G.~P.~Lepage,
			%www.github.com/gplepage/lsqfit (2014)


%%\cite{1112.5642}
%\bibitem{1112.5642} 
  %C.~M.~Bouchard {\it et al.} [Fermilab Lattice and MILC Collaborations],
  %%``Neutral $B$ mixing from $2+1$ flavor lattice-QCD: the Standard Model and beyond,''
  %%PoS LATTICE {\bf 2011}, 274 (2011)
	%\pos{PoS(Lattice 2011)274} (2011)
  %[arXiv:1112.5642 [hep-lat]].
  %%%CITATION = ARXIV:1112.5642;%%
  %%23 citations counted in INSPIRE as of 24 Nov 2013

	
\bibitem{1403.6796}
	A.~Bazavov {\it et al.} [Fermilab Lattice and MILC Collaborations],
	\pos{PoS(LATTICE2013)394} (2013)
	[arXiv:1403.6796 [hep-lat]].

\bibitem{1407.3772}
	A.~Bazavov {\it et al.} [Fermilab Lattice and MILC Collaborations],
    Phys.\ Rev.\ D {\bf 90}, 074509 (2014)
	[arXiv:1407.3772 [hep-lat]].

	
\end{thebibliography}
\end{document}